\documentclass{PoS}
\def\laeq{\raise.2ex\hbox{$<$}\kern-.75em\lower.9ex\hbox{$\sim$}\,}
\def\gaeq{\raise.2ex\hbox{$>$}\kern-.75em\lower.9ex\hbox{$\sim$}\,}

\title{Geometric modelling of radio and $\gamma$-ray light curves of 6 \textit{Fermi} LAT pulsars}
\ShortTitle{Geometric modeling of radio and $\gamma$-ray light curves of 6 \textit{Fermi} LAT pulsars}
\author{\speaker{A.~S. Seyffert}$^{a,b}$, C. Venter$^a$, O.~C. de Jager$^a$, and A.~K. Harding$^b$\\
\llap{$^a$} Centre for Space Research, North-West University, Potchefstroom Campus, 2520 Potchefstroom, South Africa\\
\llap{$^b$} E-mail: \email{20126999@nwu.ac.za}\\
\llap{$^c$} NASA Goddard Space Flight Center, Greenbelt, MD 20771, USA\\
}
\dedicated{In loving memory of Okkie de Jager.}

\abstract{The \textit{Fermi} Large Area Telescope (LAT) has recently reported the detection of pulsed $\gamma$-rays from~6 young pulsars (J0631+1036, J0659+1414, J0742$-$2822, J1420$-$6048, J1509$-$5850, and J1718$-$3825), all exhibiting single-peaked pulse profiles \cite{Weltevrede10}. High-quality radio polarization data are also available for~5 of these pulsars, allowing derivation of constraints on their viewing geometries. We obtain independent constraints on the viewing geometries of these pulsars by using a geometric pulsar emission code to model the \textit{Fermi} LAT and radio light curves. We find fits for the magnetic inclination and observer angles $\alpha$ and $\zeta$ with typical errors of $\sim5^\circ$. Our results are generally consistent with those obtained by \cite{Weltevrede10}, although we do find differences in some cases. Our model may lastly provide a framework to constrain the radio emission altitude.}

\FullConference{25th Texas Symposium on Relativistic Astrophysics\\
         December 6-10, 2010\\
         Heidelberg, Germany}

\begin{document}
\section{Introduction}
  The Large Area Telescope (LAT) is the primary instrument onboard the \textit{Fermi Gamma-ray Space Telescope} which was launched on 11 June 2008. \textit{Fermi} LAT is an imaging, wide-field-of-view, high-energy $\gamma$-ray telescope covering the energy range from 20 MeV to >300 GeV. \textit{Fermi} is currently performing an all-sky survey, taking advantage of its large field-of-view and high sensitivity to scan the entire sky every three hours \cite{Atwood09}. One of the important products to date of this all-sky survey is the first \textit{Fermi} LAT catalogue of $\gamma$-ray pulsars \cite{Abdo09_Cat} produced using the first six months of \textit{Fermi} LAT data, and increasing the number of known $\gamma$-ray pulsars from at least~6 \cite{Thompson01} to 46. The number of $\gamma$-ray pulsars continues to increase and is approaching 100 \cite{Guillemot11}. This dramatic increase offers the first opportunity to study a sizable population of these high-energy objects.

  We study~6 pulsars (J0631+1036, J0659+1414, J0742$-$2822, J1420$-$6048, J1509$-$5850, and J1718$-$3825) detected by \textit{Fermi}, all exhibiting single-peaked pulse profiles at energies > 0.1~GeV within current statistics. These are also detected in the radio band. Both of these properties aid in constraining the possible solution space for the respective pulsar geometries. 
  A study of the pulsars' geometric parameters, using the rotating vector model (RVM, \cite{Radhakrishnan69}) for radio polarization, as well as a model describing the phase difference between the polarization angle and pulsar light curve centres \cite{Blaskiewicz91}, has been performed \cite{Weltevrede10}. They constrained the solution space for the geometries of the individual pulsars using RVM fits to the radio polarization data, as well as predictions for the value of the half opening angle $\rho$ of the radio beam derived from the radio pulse width (e.g., \cite{Gil84}). The aim of this study is to obtain similar constraints on the magnetic inclination and observer angles ($\alpha$ and $\zeta$) of these pulsars using an independent, multiwavelength approach. We use a geometric pulsar emission code (e.g., \cite{Venter09}) to model the \textit{Fermi} LAT and radio light curves, and fit the predicted radio and $\gamma$-ray profiles to the data concurrently, thereby significantly constraining $\alpha$ and $\zeta$. This also allows us to test the consistency of the various approaches used to infer the pulsar geometry.

\section{Model}
  We use an idealized picture of the pulsar system, wherein the magnetic field has a retarded dipole structure \cite{Deutsch55} and the emission originates in regions of the magnetosphere (refered to as `gaps') where the local charge density is sufficiently lower than the Goldreich-Julian charge density \cite{Goldreich69}. These gaps facilitate particle acceleration and radiation. We assume that there are constant-emissivity emission layers embedded within the gaps in the pulsar's co-rotating frame. The location of these emission layers determines the shape of the light curves, and there are multiple models for the geometry of the magnetosphere describing different possible gap configurations.

  We included two models for the $\gamma$-ray emission regions in this study, namely the Outer Gap (OG) and Two-Pole Caustic (TPC) models. In the both the OG and TPC models (the Slot Gap model \cite{Arons83} may serve as physical basis for the latter), emission originates from narrow gaps along the last open field lines. In the OG model \cite{Cheng86}, emission originates above the null charge surface and interior to the last open field line, while the TPC gap starts at the stellar surface. We assume an OG emission layer position spanning $r_{\rm OVC}=0.90$ to $0.95$, while we set $r_{\rm OVC}=0.95$ to $1.00$ for the TPC case (with $r_{\rm OVC}$ the open volume coordinate which is similar to the polar angle normalized with respect to the polar cap angle \cite{Dyks04}). The special relativistic effects of aberration and time-of-flight delays, which become important in regions far from the stellar surface (especially near the light cylinder), together with the curvature of the magnetic field lines, cause the radiation to form caustics (accumulated emission in narrow phase bands) which are detected as peaks in the observed light curve \cite{Dyks03,Morini83}.

   We used the empirical radio cone model of \cite{Story07}, where the beam diameter, width, and altitude are functions of the pulsar period $P$, its time derivative $\dot{P}$, and the frequency of observation $\nu$, in order to obtain predictions for the radio light curves. (This is different from the approach taken by \cite{Weltevrede10}, as we have a different prescription for the radio emission altitude and cone half opening angle, and we do not use the polarization data.)

\section{Results}
\begin{figure}
   \includegraphics[width=\textwidth]{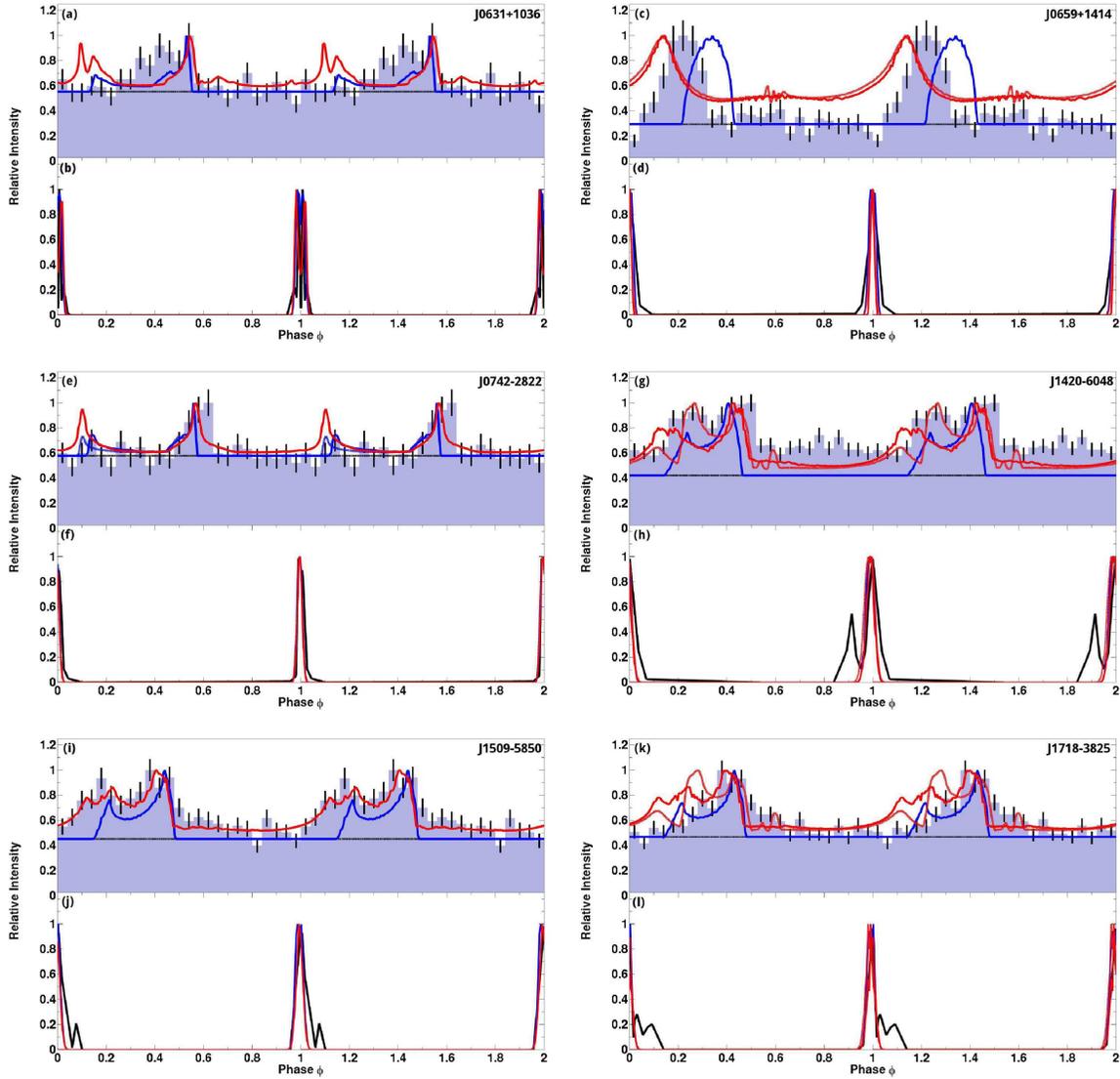}
   \caption{Model light curve fits to the $\gamma$-ray (panels a, c, e, g, i, and k) and radio (panels b, d, f, h, j, and l) data for each of the pulsars. Blue lines indicate OG fits, red lines TPC ones; dashed lines indicate alternative fits (see Table~1). The radio curves result from a conal radio model, but using the same geometry as the $\gamma$-ray curves. The histograms are \textit{Fermi} data, while the black lines are 1.4~GHz radio data (reproduced from \cite{Weltevrede10}).}\label{fig1}
 \end{figure}

\begin{figure}
   \includegraphics[width=\textwidth]{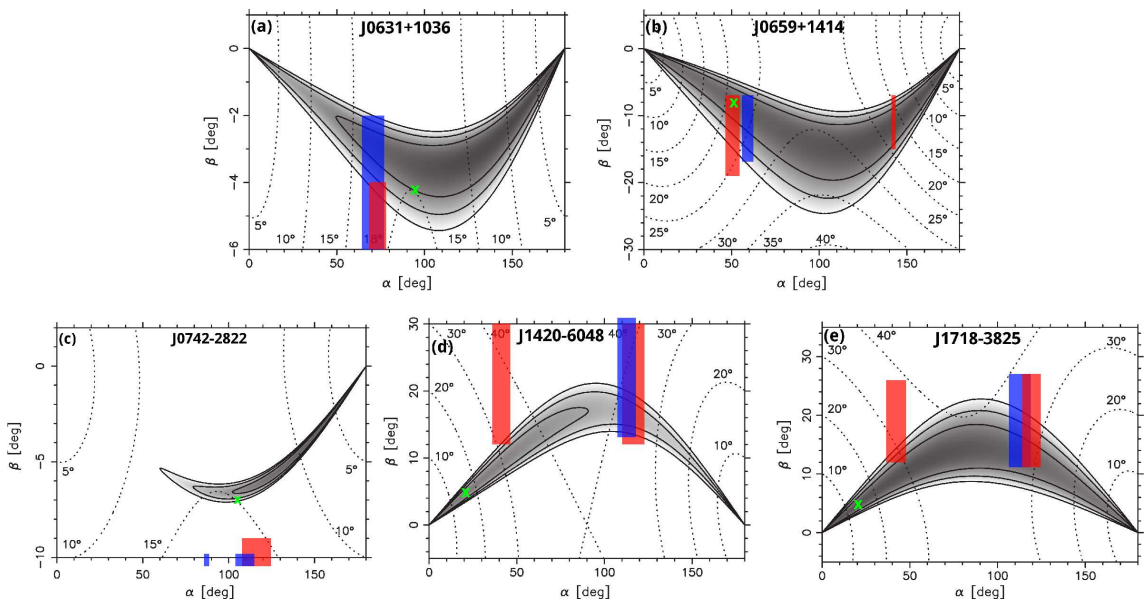}
   \caption{Independent model constraints on $\alpha$ and $\beta = \zeta-\alpha$ for each of the pulsars. Black contours are reproduced from \cite{Weltevrede10}, the green cross signifies their optimal solution. The blue areas are for the OG model, and red for the TPC model (solution contours from this work). Contours extending beyond the $\beta$-axis indicate that our solutions are not contained in the plotted area.}\label{fig2}
 \end{figure}

  Figure~\ref{fig1} indicates model light curve fits to each of the pulsars, for both the OG and TPC models. Figure~\ref{fig2} indicates significance contours in $\alpha-\beta$ space from \cite{Weltevrede10} resulting from radio polarization modelling, as well as constraints obtained from the current study. We infer values for $\alpha$ and $\zeta$ for each pulsar with typical errors of $\sim5^\circ$ (see Table~\ref{tab}). The latter have been estimated by comparing predicted light curves for different $(\alpha,\zeta)$ combinations for a $1^\circ$ resolution, and constraining the optimal solution space by eye. The tabulated values are the average values of $\alpha$, $\zeta$ and $\beta$ implied by the solution contours, while the errors are chosen conservatively so as to include the full (non-rectangular) contour. We note that once a solution is found at a particular $\alpha$ and $\zeta$, it is worthwhile to study the light curve at the position $\alpha^\prime=\zeta$ and $\zeta^\prime=\alpha$ (i.e., when the angles are interchanged). Such a complementary solution may provide a good fit in some cases (see the alternative solutions listed in Table~\ref{tab}), due to the symmetry of the model (i.e., constant beam emissivity assumed) as well the symmetry in pulsar geometry (i.e., $|\beta|$ remaining almost constant under this transformation).

\begin{table}[b]
\begin{center}
  \begin{tabular}{|c|c|c|c|c|c|c|c|}
\hline
& \multicolumn{2}{|c|}{J0631+1036} & \multicolumn{3}{|c|}{J0659+1414} & \multicolumn{2}{|c|}{J1509$-$5850}\\
& OG & TPC & OG & TPC & TPC$^1$ & OG & TPC\\
\hline
$\alpha$ & $74\pm5$  & $71\pm6$ & $59\pm3$  & $50\pm4$  & $38\pm1$ & $66\pm4$  & $61\pm5$\\
$\zeta$  & $67\pm4$  & $66\pm7$ & $48\pm3$  & $39\pm4$  & $50\pm4$ & $50\pm7$  & $44\pm7$\\
$\beta$  & $-6\pm2$  & $-5\pm3$ & $-12\pm5$ & $-13\pm6$ & $11\pm4$ & $-18\pm8$ & $-18\pm8$\\
\hline
\end{tabular}
\vskip0.3cm
\begin{tabular}{|c|c|c|c|c|c|c|c|c|c|}
\hline
& \multicolumn{3}{|c|}{J0742$-$2822} & \multicolumn{3}{|c|}{J1420$-$6048} & \multicolumn{3}{|c|}{J1718$-$3825}\\
& OG & OG$^1$ & TPC & OG & TPC & TPC$^1$ & OG & TPC & TPC$^1$\\
\hline
$\alpha$ & $86\pm3$   & $71\pm6$ & $64\pm8$  & $67\pm5$  & $64\pm6$ & $42\pm5$ & $67\pm6$   & $61\pm5$  & $42\pm6$\\
$\zeta$  & $71\pm5$   & $86\pm4$ & $80\pm4$  & $45\pm7$  & $43\pm8$ & $63\pm5$ & $48\pm6$   & $43\pm6$  & $62\pm5$\\
$\beta$  & $-16\pm6$  & $16\pm6$ & $15\pm6$ & $-22\pm9$  & $-21\pm9$ & $21\pm9$ & $-19\pm8$ & $-19\pm8$ & $19\pm7$\\
\hline
  \end{tabular}
\end{center}
\caption{Best-fit results for the modeled pulsar light curves (all angles are in degrees).}\label{tab}
$^1$ Alternative best-fit solution (upon interchange of $\alpha$ and $\zeta$).
\end{table}

\section{Conclusions}
  Generally, we find consistent results compared to those of \cite{Weltevrede10}, and our much smaller solution contours usually overlap with their best-fit contours, although we do find different geometeries for some pulsars. For better comparison, it should be noted that our model has the symmetric property (due to the assumption that the emission is symmetric for both magnetic pole) that when one transforms either $\alpha^\prime = 180^\circ - \alpha$ or $\zeta^\prime = 180^\circ - \zeta$, the light curve remains identical, but only shifts $180^\circ$ in phase, as one is then viewing the opposite magnetic pole. This shift may be neglected due to the fact that the light curves are aligned with the radio zero as defined by the data (radio and $\gamma$-ray curves are translated in unison so as to preserve the predicted radio-to-$\gamma$ lag). When doing both of these transformations, there is no shift in the light curve, but $\beta^\prime$ becomes $-\beta$. 

  It is noteworthy that our geometric modelling is able to constrain $\alpha$ significantly better compared to the constraining power of the RVM. Furthermore, given the small range of $\beta$ compared to the large range of $\alpha$ encompassed by the contours of \cite{Weltevrede10}, it is reassuring to note that we reproduce almost all of the $\beta$-values within errors. The strength of the constraints we are able to derive on $\alpha$ and $\zeta$ (or $\beta$) derives from the combination of having to reproduce the $\gamma$-ray profile shape, the radio peak multiplicity, as well as the radio-to-$\gamma$ phase lag. Furthermore, the fact that these pulsars have single peaks implies an even smaller possible solution space than for double-peaked profiles, which are more common in these $\gamma$-ray models \cite{Watters09}.

  Both the OG and TPC models provide reasonable light curve fits, neither being strongly preferred. The OG sometimes predict peaks that are too narrow, double peaks with bridge emission (between the two peaks) which is too low to reproduce the data, and no off-peak emission. On the other hand, the TPC sometimes overpredicts the background, in some cases predicts an earlier peak which is not found in the data (double-peaked profile), and generally predicts relatively broader peaks. The latter model also exhibits greater variety of peak shapes, and fills a larger region of phase space.

  We find that for each pulsar the impact angle $\beta$ remains the same between the OG and TPC geometries, even if the inferred $\alpha$ and $\zeta$ differ in each case. Also, $\beta$ is quite small ($\laeq20^\circ$). This is due to the constraints imposed by the $\gamma$-ray and radio pulse shapes (as well as their being visible in the first place). It is quite striking that $\beta<0$ in most cases, so that the observer samples radiation coming from above the magnetic axis (i.e., $\zeta < \alpha$). Interestingly, the OG-predicted $\alpha$ generally seems to be larger than that predicted by the TPC geometry. This results from the fact that the OG emission layer lies closer to the magnetic axis than that of the TPC model, so to obtain the correct radio-to-$\gamma$ lag, the $\alpha$ must be a little larger. Lastly, we also find that reproduction of the radio-to-$\gamma$ phase lag is sensitive to the value of $\zeta$.

  Our approach may provide constraints on the radio emission geometry for the different pulsars. For example, a much better OG (TPC) fit may be obtained for PSR J0659+1414 if the radio cone's altitude is lowered (increased), as this will decrease (increase) the radio-to-$\gamma$ phase lag due to a smaller (larger) aberration effect. (Alternatively, changing the position of the emission layer will also lead to changes in the $\gamma$-ray peak phase.) Furthermore, the observed radio profile width may be used to constrain the cone width for a given emission height.

  Future work includes applying a Monte Carlo Markov Chain approach for finding optimal light curve fits in multivariate phase space \cite{Johnson11} for these pulsars, constraining the radio beam properties, and comparing the new best-fit solutions (using more \textit{Fermi} data) obtained from the different approaches.

\section*{Acknowledgements}                                                                                   
This work is based upon research supported by the South African Research Chairs Initiative of the Department of Science and Technology and National Research Foundation. A.K.H. acknowledges support from the NASA Astrophysics Theory Program.


\begin{thebibliography}{99}
\bibitem[1]{Abdo09_Cat} Abdo, A.~A. et al. 2010, ApJS, 187, 460
\bibitem[2]{Arons83} Arons, J. 1983, ApJ, 266, 215
\bibitem[3]{Atwood09} Atwood, W.~B. et al. 2009, ApJ, 697, 1071
\bibitem[4]{Blaskiewicz91} Blaskiewicz, M., Cordes, J.~M., \& Wasserman, I. 1991, ApJ, 370, 643
\bibitem[5]{Cheng86} Cheng, K. S., Ho, C., \& Ruderman, M. 1986, ApJ, 300, 522
\bibitem[6]{Deutsch55} Deutsch, A. J. 1955, Annales d'Astrophysique, 18, 1
\bibitem[7]{Dyks03} Dyks, J., \& Rudak, B. 2003, ApJ, 598, 1201
\bibitem[8]{Dyks04} Dyks, J., Harding, A.~K., \& Rudak, B. 2004, ApJ, 606, 1125
\bibitem[9] {Gil84} Gil, J.~A., Gronkowski, P., \& Rudnicki, W. 1984, A\&A, 132, 312
\bibitem[10]{Goldreich69} Goldreich, P., \& Julian, W.~H. 1969, ApJ, 157
\bibitem[11]{Guillemot11} Guillemot, L., these proceedings
\bibitem[12]{Johnson11} Johnson, T.~J., Harding, A.~K., \& Venter, C. 2011, in preparation.
\bibitem[13]{Morini83} Morini, M. 1983, MNRAS, 202, 495
\bibitem[14]{Radhakrishnan69} Radhakrishnan, V., \& Cooke, D.~J. 1969, ApJL, 3, 225
\bibitem[15]{Story07} Story, S. A. et al. 2007, ApJ, 671, 713
\bibitem[16]{Thompson01} Thompson, D.~J. 2001, in AIP Conf. Ser., ed. F.~A.~Aharonian \& H.~J.~V{\"o}lk, 103 (astro-ph/0101039)
\bibitem[17]{Venter09} Venter, C., Harding, A.~K., \& Guillemot, L. 2009, ApJ, 707, 800
\bibitem[18]{Watters09} Watters, K.~P., Romani, R.~W., Weltevrede, P., \& Johnston, S. 2009, ApJ, 695, 1289
\bibitem[19]{Weltevrede10} Weltevrede, P. et al. 2010, ApJ, 708, 1426
\end{thebibliography}
\end{document}